\documentstyle[12pt,epsf]{article}
\newcommand{\bRe}{{\rm Re}}
\newcommand{\bIm}{{\rm Im}}
\newcommand{\btr}{{\rm tr}}
\newcommand{\bu}{{\bf U}}
\newcommand{\bR}{{\bf R}}
\newcommand{\bZ}{{\bf Z}}
\newcommand{\bC}{{\bf C}}
\newcommand{\bPC}{{\bf CP}}

\newtheorem{lemma}{Lemma}[section]

\begin{document}
\title{
\rightline{\small  HUB--EP--99/01}
\vspace{1cm}
Definition of Magnetic Monopole Numbers for $SU(N)$ Lattice
Gauge-Higgs Models
}
\author{
S. Hollands$^1$\thanks{Supported by 
        Graduiertenkolleg `Strukturuntersuchungen, 
        Pr\"azisionstests und Erweiterungen des Standardmodells der 
        Elementarteilchenphysik', DFG-GRK 271/1-96} 
        and
M. M\"uller-Preussker$^2$,  \\
{\small{\it $^1$ Department of Mathematics,
             University of York, UK}} \\
{\small{\it $^2$ Institut f\"ur Physik,
             Humboldt-Universit\"at zu Berlin, Germany}}
}
\date{December 29, 1998}


\maketitle

\begin{abstract}
A geometric definition for a magnetic charge of Abelian 
monopoles in $SU(N)$ lattice gauge theories with Higgs fields
is presented.  The corresponding local monopole number defined 
for almost all field configurations does not require gauge fixing
and is stable against small perturbations.  Its topological content 
is that of a 3-cochain. A detailed prescription for calculating 
the local monopole number is worked out.

Our method generalizes a magnetic charge definition previously invented 
by Phillips and Stone for $SU(2)$. 
\end{abstract}

\newpage

\section{Introduction}
The investigation of the complicated vacuum structure of quantized
Yang-Mills theories requires the use of non-perturbative methods.
The lattice regularization is the most convenient
in this respect.  Among other applications it allows to identify those
excitations (monopoles, instantons, etc.) which are expected to
dominate the Euclidean path integral at large scales and to explain
important features like chiral symmetry breaking and quark confinement
in QCD.

Monopoles in which we take interest in this paper
form the basis for the dual superconductor scenario of confinement
\cite{mandelst, hooft}. According to this scenario the condensation
of (anti-) monopoles causes the color-electric flux to be squeezed
into string-like flux tubes (dual Meissner effect). The consequence
is a linearly rising potential between static quarks.

Whether extended non-Abelian 't Hooft-Polyakov-like monopoles
(see \cite{Smit_vdS}) or singular Abelian Dirac-like monopoles play
the major role in the confinement phenomenon has not been settled yet.
The majority of investigations concentrates on the latter kind
of excitations. These can be identified by Abelian projection
after an appropriate gauge has been fixed \cite{KSW}.
Up till now the maximally Abelian gauge was considered to be
the most favourable gauge \cite{suzuki,BIMZM,BMM}
(for a recent review see e.g. \cite{gbali}). For this gauge the
resulting $~U(1)~$ degrees of freedom as well as the corresponding
Abelian monopoles were shown to reproduce almost completely the string
tension of the full theory (so-called Abelian and monopole dominance,
respectively) \cite{shiba,bali}.

Other gauges -- defined for instance by diagonalizing gauge observables
as parallel transporters along closed paths (e.g. the Polyakov loop)
which transform like Higgs fields in the adjoint representation --
have been studied, too \cite{reinh_wipf}.
Based on appropriately defined order parameters according to a
construction of Fr\"ohlich and Marchetti \cite{fromarch}
the condensation of Abelian monopoles in the confinement phase
was demonstrated to occur widely independent of the gauge chosen
\cite{DGPCV}.

But a general gauge independent proof for the validity of the dual
superconductor picture is still missing.

Therefore, it would be worthwhile to have a gauge-independent
and -- at the same time -- geometric lattice definition of the monopole
charge. This, in general, requires the existence of a Higgs field in the
adjoint representation. Magnetic fluxes and the magnetic charge,
respectively, can easily be obtained by projecting the non-Abelian
plaquette variables onto the local colour direction of the
Higgs field. However, the naive discretization does not
provide any geometric meaning of the resulting charge. Nevertheless,
corresponding investigations of magnetic fluxes have provided
useful insight into the phase structure of the 4d $SU(2)$
Georgi-Glashow model \cite{BIMMZ}.

A first geometrical definition for the monopole number has been given
in \cite{phsto1,phsto2}. The lattice gauge field together with the Higgs
field determines local maps from the boundary of cubes into the coset
space $SU(2)/U(1)$. The winding numbers of these maps provide the
local integer-valued monopole charge, for each cube, or alternatively
speaking, a closed magnetic current on the dual lattice. In our present
work we are going to generalize the definition to a $SU(N)$ gauge-Higgs
theory. The Higgs fields employed are considered to be $\bPC^{N-1}$-valued
fields. We shall present all the details necessary for developing a computer
code for the calculation of the winding numbers from the local lattice
field data.

The resulting magnetic charge definition can be employed for explorations
of the phase structure of $SU(N)$ gauge-Higgs models. But,
it might be worthwhile also for the application to pure $SU(N)$ gauge
theories, when an auxiliary Higgs field is determined as proposed for
the recently invented Laplacian gauge fixing \cite{AvS}.

Recently, in the context of investigations of the electro-weak phase
transition for the $SU(2)$ theory another version of an integer-valued,
gauge independent lattice magnetic charge definition was invented
\cite{ChBI}. It remains an interesting task to work out the interrelation
between the different definitions, if there is any.

Section 2 will provide the general description and the necessary notations.
In Section 3 we present the computation of the monopole number. Some
detailed proofs are shifted into the Appendix. In Section 4 we shall draw
the conclusions.

\section{Description of the model and notation}
In this work we want to consider a model consisting of an $SU(N)$
lattice gauge field $\bu = \{U_{ij}\}$ coupled to a 
Higgs field $\Phi = \{\phi_i\}$ transforming with respect to the fundamental
representation. $i, j, \dots$ denote the lattice sites and 
pairs $\langle ij \rangle$ the corresponding links of a 4-dimensional 
simplicial lattice. Without loss of generality for our construction of
the local monopole number we work with a Higgs field of fixed length, 
$\| \phi_i \| = 1$. In the following we denote by $U_{ijk \dots } = 
U_{ij}U_{jk} \dots$ parallel transporters longer than one link.

As the phase of the vector $\phi_i \in \bC^N$ will be irrelevant for the
monopole charge, it is sufficient to identify the corresponding equivalence 
class $[\phi_i]$. The latter can be represented by the $N \times N$ projection 
matrix $\Phi_i = \phi_i \otimes \phi_i^*$ projecting on the subspace 
spanned by the vector $\phi_i$. In this sense the Higgs field takes
values in a $\bPC^{N-1}$ manifold. 

The action of a complex invertible matrix $V$ on elements of $\bPC^{N-1}$ 
will be denoted by `$*$', e.g. $V*\Phi$ means the projector
$V\phi \otimes (V\phi)^*/\|V\phi\|^2$ or the class $[V\phi]$.
For the lattice we assume an ordering of the 
lattice sites and denote the different kinds of simplices 
(sites, links, triangles, tetrahedra, 
4-tetrahedra) by ordered tuples of vertices such as 
$\langle ij \dots k \rangle$. If we consider a site $i$ of a given simplex 
$\langle ij \dots k \rangle$ as the coordinate origin,  a cube 
$c_{\langle ij \dots k\rangle}$ parametrized by coordinates 
$0 \le s_j, \dots, s_k \le 1$ can be defined.

\section{Definition of the local monopole number}
Our aim in this section is to define for a given tetrahedron
$\Delta = \langle ijkl \rangle$ a map $X_\Delta$ from the boundary
of the 3-dimensional cube $c_{\langle ijkl \rangle}$ to $\bPC^{N-1}$. 
This map is constructed from the link and Higgs variables at the 
particular tetrahedron $\Delta$. 

Since $\Pi_2(\bPC^{N-1}) \cong \bZ$, the homotopy class of such a map
is represented by an integer $n_\Delta$. This integer will be referred
to as the local monopole number at the particular simplex. It will 
be shown to have the following properties:
\begin{enumerate}
\item[(a)] {\bf Gauge invariance:} $n_\Delta$ is independent of a particular
choice of gauge.

\item[(b)] {\bf Local conservation:}\\
The assignment $\Delta \mapsto
n_\Delta$ represents a cochain in the sense that 
$$ (\delta n)_T = \sum_{\Delta \in \partial T} (-)_\Delta n_\Delta = 0, $$
the sign depending on the relative orientation of $\Delta$ in the
boundary of a $4$-tetrahedron $T$. If $\mu$ is the link dual to 
$\Delta$ we may equivalently say that $M_\mu = n_\Delta$ is a conserved
(divergence free) magnetic current on the dual lattice.

\item[(c)] {\bf Geometric significance:}\\
The collection of the $X_\Delta$'s 
define a section of a $SU(N)/U(N-1)$-bundle over the $2$-skeleton
of the lattice. This section may be regarded as an interpolation of
the Higgs field which is as horizontal as possible with respect to 
the lattice gauge field.
\end{enumerate}  

The maps $X_\Delta$ are worked out stepwise for higher simplices. In order
to state their relevant properties, we need $GL(N)$-valued parallel 
transport functions $V_{\langle ij \dots kl \rangle}(s_j, \dots, s_k)$ 
as introduced in \cite{phsto2}. These functions essentially measure to 
what extent the lattice gauge field fails to be flat. They are
an important ingredient in the construction of $SU(N)$ bundles
associated to lattice gauge fields. In our case, the relevant formulae 
read
\begin{eqnarray*}
V_{\langle ij \rangle}  &=& U_{ij}, \\ 
V_{\langle ijk \rangle} &=& s_j U_{ijk} + (1 - s_j)U_{ik}, \\ 
V_{\langle ijkl \rangle}&=& s_j s_k U_{ijkl} + (1 - s_j)s_k U_{ikl}
                      +s_j(1 - s_k) U_{ijl} + (1 - s_j)(1 - s_k) U_{il}, 
\end{eqnarray*}
where $0 \le s_j, s_k \le 1$. Let us assume for simplicity that the
simplex $\Delta$ in question is $\langle 01 \dots k \rangle$. Then
$X_\Delta$ is assumed to meet the following requirements:
\begin{eqnarray}
X_{\langle 0 \dots k \rangle}(\dots, s_r = 1, \dots)&=&
V_{\langle 0 \dots r \rangle}(s_1, \dots)* 
X_{\langle r  \dots k \rangle}(s_r, \dots),\label{1}\\ 
X_{\langle 0 \dots k \rangle}(\dots, s_r = 0, \dots)&=& 
X_{\langle 0 \dots \slash\!\!\!r \dots k \rangle}
(s_1, \dots, \slash\!\!\!s_r, \dots)\label{2}, 
\end{eqnarray}
and likewise for all other simplices, with the obvious replacements.
It follows from the results in \cite{phsto2} that such a collection
of maps defines a section over the $2$-skeleton in a bundle with 
principal fiber $\bPC^{N-1}$. It is also seen that the local winding 
number of this section at a tetrahedron $\Delta = \langle ijkl \rangle$ 
is just the winding number of $X_{\langle ijkl \rangle}:
\partial c_{\langle ijkl \rangle} \rightarrow \bPC^{N-1}$. 
Let us now explicitly construct this map. 

Requirements (\ref{1}) and (\ref{2}) suggest to define the $X_\Delta$'s 
inductively for simplices of increasing dimension. This is done in four 
steps:

{\bf Step 0:} As the $X_\Delta$'s essentially interpolate the Higgs field, 
we set $X_{\langle i \rangle} = \Phi_i$ for the lattice sites.

{\bf Step 1:} On the far end $s_j = 1$ of the interval $c_{\langle
ij \rangle}$, we must have, by requirement (\ref{1})
\begin{eqnarray*}
X_{\langle ij \rangle}(s_j = 1) = V_{\langle ij \rangle} * 
X_{\langle j \rangle}.
\end{eqnarray*} 
On the base $s_j = 0$, $X_{\langle ij \rangle}$ must be $X_{\langle i
\rangle}$, by requirement (\ref{2}). We thus interpolate geodesically in 
$\bPC^{N-1}$ for intermediate values of $s_j$. 

{\bf Step 2:} On the far sides $s_j = 1$ or $s_k = 1$ of the square
$c_{\langle ijk \rangle}$, we must have, by requirement (\ref{1})
\begin{eqnarray*}
(A)\quad&&
X_{\langle ijk \rangle}(s_j = 1, s_k)  =  V_{\langle ij \rangle} * 
X_{\langle jk \rangle}(s_k),\\
(B)\quad&&
X_{\langle ijk \rangle}(s_j, s_k = 1)  =  V_{\langle ijk \rangle}(s_j)* 
X_{\langle k \rangle}.
\end{eqnarray*}
Again, on the base $s_j = s_k = 0$, $X_{\langle ijk \rangle}$ must be 
$X_{\langle i\rangle}$, by requirement (\ref{2}). For intermediate values
we interpolate geodesically from the the far sides to the base.

{\bf Step 3:} In the last step we proceed as before. Again, on the far 
faces  $s_j = 1, s_k = 1$ or $s_l = 1$ of the cube
$c_{\langle ijkl \rangle}$, we must have, by requirement (\ref{1})
\begin{eqnarray*}
(I)&&\quad 
X_{\langle ijkl \rangle}(s_j = 1, s_k, s_l) = V_{\langle ij \rangle} * 
X_{\langle jkl \rangle}(s_k, s_l),\\
(II)&&\quad
X_{\langle ijkl \rangle}(s_j, s_k = 1, s_l) = V_{\langle ijk \rangle}(s_j)* 
X_{\langle kl \rangle}(s_l),\\
(III)&&\quad
X_{\langle ijkl \rangle}(s_j, s_k, s_l = 1) = V_{\langle ijkl \rangle}
(s_j, s_k)*X_{\langle l \rangle}.
\end{eqnarray*}
As above, on the base $s_j = s_k = s_l = 0$, $X_{\langle ijkl \rangle}=
X_{\langle i\rangle}$, by requirement (\ref{2}). Unlike in the previous cases, 
we cannot interpolate from the the far sides to the base since there might be
an obstruction. Nevertheless, we might do so on the boundary of our
cube. This completes our construction. Requirements (\ref{1}) and 
(\ref{2}) are more or less obviously satisfied.

In the next chapter we shall see how the winding number of 
$X_{\langle ijkl \rangle}$ may be calculated. 
   
\section{Calculation of the monopole number}
\begin{figure}[!thb]
\vspace*{0.5cm}
\begin{center}
\hspace*{0.5cm}
\hbox{
\epsfxsize=9cm\epsfysize=9cm
\epsfbox{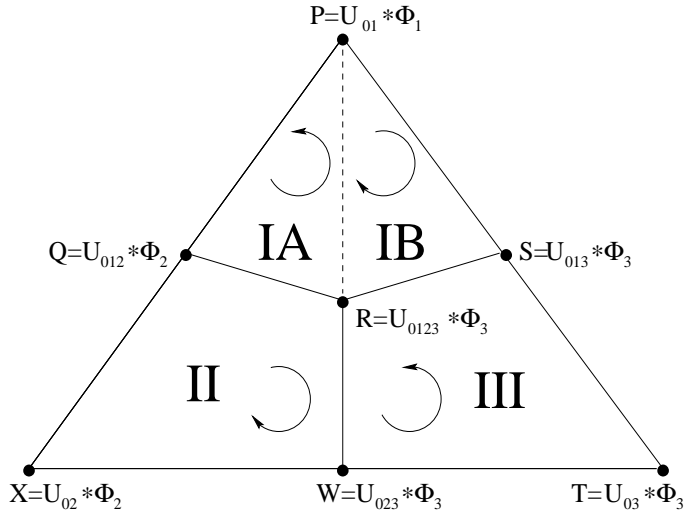}
}
\vspace*{-1.5cm}
\caption{\sl
The image of the surfaces $C^1_k$ as treated in cases
(I), (II), (III).
}
\end{center}
\label{fig1}
\end{figure}

We have to determine the class of $X_\Delta$. Take a simplex 
$\Delta = \langle 0123 \rangle$. The boundary of the unit cube 
$c_\Delta$ with base $\langle 0\rangle$ is the union of six faces 
$C^1_k, C^0_k \,\,(k = 1, 2, 3)$, where
$$ C^0_k = \{ (s_1, s_2, s_3): \quad s_k = 0 \}, \quad
   C^1_k = \{ (s_1, s_2, s_3): \quad s_k = 1 \}. $$
The monopole number may be computed as an intersection number
\cite{chern},
\begin{eqnarray*}
n_\Delta = X_\Delta(\partial c_\Delta) : E^\perp,
\end{eqnarray*}
$E^\perp$ denoting the $(2N-4)$-dimensional space of all projectors
perpendicular to some rank one projector $E$. The intersection number 
does not depend on the particular projector $E$ chosen. To facilitate the 
calculations, we take it to be $\Phi_0$.
The strategy is to compute the intersection number separately for the 
six different faces of the
surface of our cube. As geodesic interpolation is needed for the
definition of $X_\Delta$, let us collect some facts about the geometry of
complex projective spaces (see also \cite{UB,TH,Helg}). 
Two points $P, Q$ in the complex
projective space may be joined by a unique shortest geodesic if and
only if $PQ \neq 0$. Furthermore, for such points it holds:
\begin{lemma} \label{33}
The unique shortest geodesic between two points $P, Q$ intersects
$E^\perp$ if and only if $\bRe\,\btr [PQE] \le 0$ and $\bIm\,\btr [PQE] = 0$.
Moreover, such an arc can intersect $E^\perp$ only once.
\end{lemma}
We shall also need the following fact about triangles.
\begin{lemma} \label{22}
Let $P, Q, R$ be three points such that there is a unique shortest geodesic
between any two of them. Let $\triangle$ be the surface spanned by all 
geodesics between $P$ and the geodesic arc $\widehat{QR}$. This surface
intersects $E^\perp$ if and only if the equation $\bIm\,\varphi(t) = 0$ has
a solution $0 \le t \le 1$ satisfying $\bRe\,\varphi(t) \le 0$. Here
\begin{eqnarray}\label{400}
\varphi(t) &=& t^2\,\btr [QR]^{1/2} \btr [ERP] + \\
&&(1-t)^2\,\btr [QR]^{1/2} \btr [EQP] + t(1-t)\, 
\btr [E(QR + RQ)P].\nonumber
\end{eqnarray}
The sign of such an intersection is given by
\begin{eqnarray*}
{\rm sign\,\, Im} \left[
\frac{\partial \varphi(t)}{\partial t} \right].
\end{eqnarray*}
\end{lemma} 
Lemma \ref{33} and \ref{22} are proven in the appendix. Let us first turn 
our attention to the faces $C^0_k$. With our particular choice of 
$E$ the following lemma holds.
\begin{lemma} \label{77}
Except for a measure zero set of configurations
$(\bu, \Phi)$, the surfaces $X_{\langle 0123 \rangle}(C^0_k)$ do
not intersect with $\Phi_0^\perp$.
\end{lemma}
The proof of this lemma is given in the appendix. It follows that we only
have to consider the faces $C^1_1, C^1_2, C^1_3$, schematicly shown in
Fig. \ref{fig1} as quadrilaterals (I), (II), (III) respectively. On these faces, 
the map $X_{\langle 0123 \rangle}$ has the form given in 
Eqs.~(I), (II), (III). The monopole number will be the sum of 
the intersection numbers obtained in each case. 

\begin{figure}[!thb]
\begin{center}
\hspace*{1.0cm}
\vspace*{0.5cm}
\hbox{
\epsfxsize=5cm\epsfysize=5cm
\epsfbox{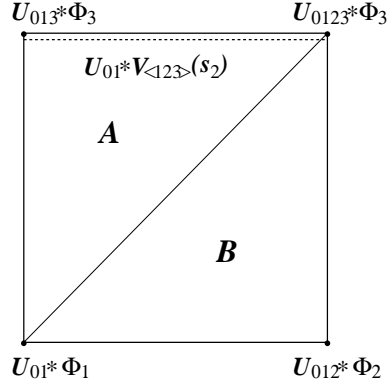} 
}
\end{center}
\caption{\sl 
The image of $C^1_1$ according to Eq. (I).
}
\vspace*{0.5cm}
\label{fig2}
\end{figure}

{\bf Case (I):} 
The image of $C^1_1$ is described by Eq. (I). As illustrated in 
Fig. \ref{fig2}, it is naturally divided 
into two triangles $\bf A, B$, according to
the structure of Eqs. (A) and (B). Let us first consider triangle {\bf A}. 
The corners of this triangle are the points $P = U_{01}*\Phi_1, 
Q = U_{012}*\Phi_2, R = U_{0123}*\Phi_3$. The surface enclosed by these 
three points is swept out by the set of all geodesics between $P$ and the 
geodesic arc $\widehat{QR}$, its orientation being fixed by the order 
$P, Q, R$ of the vertices, see Fig. \ref{fig1}. This is the situation 
described in Lemma \ref{22}, except for degenerate cases and we use this 
result to calculate the intersection number of {\bf A} with $\Phi_0^\perp$. 
Inserting the above expressions for $P, Q, R$ into Eq. (\ref{400}), 
we have to consider 
\begin{eqnarray}\label{IA}
\varphi_{IA}(t) &=& t^2 \langle \phi_0 | U_{0123}\phi_3 \rangle 
\langle U_{123}\phi_3 | \phi_1 \rangle \langle U_{01}\phi_1 | \phi_0
\rangle |\langle \phi_2 | U_{23}\phi_3 \rangle|\\ 
&+& 
(1 - t)^2 \langle \phi_0 | U_{012}\phi_2 \rangle 
\langle U_{12}\phi_2 | \phi_1 \rangle \langle U_{01}\phi_1 | \phi_0
\rangle |\langle \phi_2 | U_{23}\phi_3 \rangle|\nonumber\\
&+& t(1 - t)(\langle \phi_0 | U_{012}\phi_2 \rangle 
\langle \phi_2 |U_{23}\phi_3 \rangle 
\langle U_{123}\phi_3 | \phi_1 \rangle \langle U_{01}\phi_1 | \phi_0 \rangle
\nonumber\\
&+& \langle \phi_0 | U_{0123}\phi_3 \rangle 
\langle U_{23}\phi_3 | \phi_2 \rangle 
\langle U_{12}\phi_2 | \phi_1 \rangle \langle U_{01}\phi_1 | \phi_0 \rangle),
\nonumber
\end{eqnarray}
where the standard notation for the scalar product in $\bC^N$ has been used.

In order to find the intersections, we must look for solutions 
$0 \le t \le 1$ of $\bIm\,\varphi_{IA}(t) = 0$ for which 
$\bRe\,\varphi_{IA}(t) \le 0$. According to Lemma \ref{22}, such solutions
will contribute with the sign
\begin{eqnarray*}
{\rm sign\,\, Im} \left[
\frac{\partial \varphi_{IA}(t)}{\partial t} \right]
\end{eqnarray*}
to the intersection number. 

Triangle {\bf B} is swept out by geodesics from the dotted line 
Fig. \ref{fig2} to the point $P = U_{01}*\Phi_1$. 
This arc is parametrized by $V_{\langle 01 \rangle}
V_{\langle 123 \rangle}(s_2)*\Phi_3$. Hence by Lemma
\ref{33}, we have to find points $0 \le s_2 \le 1$ such that
$\bIm\,\varphi_{IB}(s_2) = 0$ and $\bRe\,\varphi_{IB}(s_2) 
\le 0$, where $\varphi_{IB}$ is the function
\begin{eqnarray*}
\btr \left[ 
X_{\langle 0 \rangle}
(V_{\langle 01 \rangle}*X_{\langle 1 \rangle}) 
(V_{\langle 123 \rangle}(s_2)*X_{\langle 3 \rangle})
\right].
\end{eqnarray*}
Explicitly, in terms of the parallel transporters and Higgs fields, 
this may be written as
\begin{eqnarray}\label{IB}
\varphi_{IB}(s_2) &=& s_2^2 \langle \phi_0 | U_{01}\phi_1 \rangle 
                    \langle \phi_1 | U_{123}\phi_3 \rangle 
                    \langle U_{0123}\phi_3 | \phi_0 \rangle\\
                  &+& (1 - s_2)^2 \langle \phi_0 | U_{01}\phi_1 \rangle 
                    \langle \phi_1 | U_{13}\phi_3 \rangle 
                    \langle U_{013}\phi_3 | \phi_0 \rangle\nonumber\\
                  &+& s_2(1 - s_2)\langle \phi_0 | U_{01}\phi_1 \rangle( 
                    \langle \phi_1 | U_{123}\phi_3 \rangle 
                    \langle U_{013}\phi_3 | \phi_0 \rangle\nonumber\\
                  &+& \langle \phi_1 | U_{13}\phi_3 \rangle 
                    \langle U_{0123}\phi_3 | \phi_0\rangle),\nonumber 
\end{eqnarray}
which is explicitly seen to be gauge invariant (we have suppressed 
a positive overall factor which obviously
does not affect our considerations). Values of $s_2$ with
the above properties may thus be found by elementary means. 
In the appendix, case $\alpha)$ we show that they make the following 
contribution to the intersection number, i.e. have the following sign for 
the relative orientation:
\begin{eqnarray*}
{\rm sign\,\, Im} \left[
\frac{\partial \varphi_{IB}(s_2)}{\partial s_2} \right].
\end{eqnarray*}

{\bf Case (II):} The image of $C^1_2$ is a square. 
It is obtained by `moving down' the geodesic arc $\widehat{PQ}$ to the 
arc $\widehat{XW}$, see Fig. \ref{fig1}. From the parametrization of
this surface, Eq. (II) one concludes that it will intersect the manifold
$\Phi_0^\perp$ if and only if 
\begin{eqnarray}\label{cond}
(V_{\langle 012 \rangle}(s_1)*X_{\langle 23 \rangle}(s_3))
X_{\langle 0 \rangle} &=& 0,\nonumber\\
\Longleftrightarrow (V_{\langle 210 \rangle}(s_1)*X_{\langle 0 \rangle})
X_{\langle 23 \rangle}(s_3) &=& 0,
\end{eqnarray} 
for some $0 \le s_1, s_3 \le 1$. 
From Eq. (\ref{cond}) we see that this will happen precisely if the 
geodesic arc parametrised by $X_{\langle 23 \rangle}(s_3)$ intersects
the manifold $(V_{\langle 210 \rangle}(s_1)*\Phi_0)^\perp$ for some value
$0 \le s_1 \le 1$. The arc has endpoints $X_{\langle 2 \rangle}$
and $V_{\langle 23 \rangle}*X_{\langle 3 \rangle}$. We may apply 
Lemma \ref{33} to conclude that for such an intersection we must have
$\bIm\,\varphi_{II}(s_1) = 0$ and $\bRe\,\varphi_{II}(s_1) \le 0$, where
$\varphi_{II}$ is the function 
\begin{eqnarray*}
\btr \left[ X_{\langle 2 \rangle} 
(V_{\langle 23 \rangle}*X_{\langle 3 \rangle})
(V_{\langle 210 \rangle}(s_1)*X_{\langle 0 \rangle}) \right].
\end{eqnarray*}    
Expressing everything in terms of parallel transporters and Higgs fields,
we find
\begin{eqnarray}\label{II}
\varphi_{II}(s_1) &=& s_1^2 \langle \phi_2 | U_{23}\phi_3 \rangle
\langle U_{0123}\phi_3 | \phi_0 \rangle \langle \phi_0 | U_{012}\phi_2 
\rangle\\
&+& (1 - s_1^2)^2 \langle \phi_2 | U_{23}\phi_3 \rangle
\langle U_{023}\phi_3 | \phi_0 \rangle \langle \phi_0 | U_{02}\phi_2 
\rangle\nonumber\\
&+& s_1(1 - s_1) \langle \phi_2 | U_{23}\phi_3 \rangle(
\langle U_{0123} \phi_3 | \phi_0 \rangle \langle \phi_0 | U_{02} \phi_2
\rangle\nonumber\\
&+& \langle U_{023} \phi_3 | \phi_0 \rangle 
\langle \phi_0 | U_{012} \phi_2 \rangle),\nonumber 
\end{eqnarray}
and this is directly seen to be gauge invariant. Again as in case (I), 
we have omitted an irrelevant positive overall factor. It is shown in 
the appendix, case $\beta)$ that such an intersection contributes with the 
sign
\begin{eqnarray*}
{\rm sign\,\, Im} \left[
\frac{\partial \varphi_{II}(s_1)}{\partial s_1} \right].
\end{eqnarray*}
{\bf Case (III):} From Eq. (III) we can see that the image of $C^1_3$ 
is a square, parametrized by  
$V_{\langle 0123 \rangle}(s_1, s_2)*X_{\langle 3 \rangle}$, where
$0 \le s_1, s_2 \le 1$. This square will intersect the manifold 
$\Phi_0^\perp$ precisely for values $s_1, s_2$ such that 
\begin{eqnarray*}
X_{\langle 0 \rangle}(V_{\langle 0123 \rangle}(s_1, s_2)*
X_{\langle 3 \rangle}) = 0. 
\end{eqnarray*}
We have to find such values and determine their contribution to the
intersection number. Writing everything in terms of
parallel transporters and Higgs fields, the above condition reads
$\varphi_{III}(s_1, s_2) = 0$, where
\begin{eqnarray}\label{III}
\varphi_{III}(s_1, s_2) &=& 
\langle \phi_0 \vert U_{0123} \phi_3 \rangle s_1 s_2+ 
\langle \phi_0 \vert U_{013} \phi_3 \rangle s_1 (1-s_2)\\ 
&+& \langle \phi_0 \vert U_{023} \phi_3 \rangle (1-s_1) s_2+ 
\langle \phi_0 \vert U_{03} \phi_3 \rangle (1-s_1)(1-s_2),\nonumber
\end{eqnarray}
and this expression is manifestly gauge invariant. 
Condition (\ref{III}) constitutes a system of two real equations in two 
real variables, which can
easily be solved. In the appendix, case $\gamma)$ it is demonstrated 
that any point $(s_1, s_2)$ satisfying Eq. (\ref{III}) 
contributes the value 
\begin{eqnarray*}
{\rm sign\,\, Im}
\left[
\frac{\partial \varphi_{III}(s_1, s_2)}{\partial s_1}
\frac{\partial \overline{\varphi_{III}(s_1, s_2)}}{\partial s_2} \right]
\end{eqnarray*}
to the intersection number. 

\section{Conclusions and outlook}
In this work we have given a geometric definition of local monopole
numbers for an $SU(N)$ lattice gauge-Higgs system. The definition 
yields a closed magnetic current which is stable under perturbations
of the configuration in question. It is explicitly worked out how to 
calculate the monopole number for each tetrahedron of a simplicial 
4-dimensional lattice. On the practical side, one has to solve four 
quadratic equations in one variable for each such tetrahedron. The 
coefficients of these equations are given by traces of parallel transporters 
and Higgs variables. The prescription is expicitly gauge invariant and can 
be directly implemented in a computer code. 

While the construction gives an integer local monopole number for almost all
field configurations, the question is on what kind of monopole the construction
actually triggers. In the $SU(2)$ case for the Georgi-Glashow model it has
been shown \cite{phsto1} that the construction is able to detect a discretized 
't Hooft-Polyakov monopole, provided the configuration is smooth enough. This 
question remains to be discussed for 
$N > 2$. On the other hand, for rough lattice fields as they are generated
in a Monte-Carlo simulation, one has to expect nonvanishing monopole 
charges which do not correspond to physical excitations but are
lattice artefacts (dislocations). It is then necessary to find out 
whether these configurations become sufficiently suppressed 
in the continuum limit.   

\begin{appendix}
\vspace*{1.5cm}\noindent
{\Large {\bf Appendix}}
\par\vspace*{0.5cm}\noindent
Points in $\bPC^{N-1}$ are identified with equivalence 
classes $[a] = \bC a$ (lower case letters) or projectors $A$ 
(capital letters) projecting on the subspace spanned by $a$. 
Two points $[p], [q]$ are joined by the unique 
shortest geodesic $[(1-t)\langle p|q \rangle p + t|\langle p | q 
\rangle|^2 q]$, 
provided $\langle p | q \rangle \neq 0$ (see \cite{UB}).
\par\medskip\noindent
{\it Proof of Lemma \ref{33}:} \\ 
From the form of the geodesics it is clear that the geodesic arc
$\widehat{PQ}$ will intersect $E^\perp$ iff $(1 - t)\langle e | p \rangle
\langle p | q \rangle + t |\langle p | q \rangle|^2 \langle e | q \rangle = 0$
for some $0 \le t \le 1$. This is easily seen to be equivalent to
$$ 
\bRe \big[ \langle e | p \rangle \langle p | q \rangle \langle q | e \rangle 
\big]
\le 0, \quad 
\bIm \big[ \langle e | p \rangle \langle p | q \rangle \langle q | e \rangle 
\big]= 0, 
$$   
proving the lemma. 
\par\medskip\noindent
{\it Proof of Lemma \ref{22}:} \\ 
We parametrize 
the geodesic arc $\widehat{QR}$ by $[p'(t)]$ as above. Now the triangle
$\triangle$ intersects $E^\perp$ iff some arc $\widehat{P'(t)P}$ does. 
According to lemma \ref{33} this can happen if and only if 
$\bIm\,\btr[P P'(t) E] = 0$ and $\bRe\,\btr[P P'(t) E] \le 0$. Inserting
the expression for $P'(t)$, we see that these are just the conditions
given in the lemma. The statement for the sign of the intersection is
obtained in case $\alpha.A)$ below.  
\par\medskip\noindent
{\it Proof of Lemma \ref{77}:}  \\
The image under the map $X_{\langle 
0123 \rangle}$ of the faces $C^0_k$ is obtained by coning the boundary 
of Fig. \ref{fig1} to the point $E := \Phi_0$. Let $A$ be any point in
this boundary. In order that the geodesic arc $\widehat{AE}$ intersects
$E^\perp$, it is necessary that ${\rm Re\,tr\,} AE \le 0$, by Lemma 
\ref{33}. But this is easily seen to imply $AE = 0$, i.e. $A \in
E^\perp$. So the image of the faces $C^0_k$ intersects $E^\perp$ if 
and only if the boundary of the triangle in Fig. \ref{fig2} does. On 
dimensional grounds this can happen only for a measure zero set of 
configurations.
\par\medskip\noindent
We now prove the various claims about intersection numbers made above. Let
us briefly recall how the relative orientation of two oriented, embedded 
transversally intersecting manifolds is defined. Suppose $X, Y \subset Z$ 
are oriented embedded manifolds, intersecting transversally at a point $p$.
The union of two oriented frames in $T_p X$ and $T_p Y$ then gives a
frame in $T_p Z$ whose orientation may be compared to the orientation of
$Z$. The sign $X:Y$ is defined to be $\pm 1$ according to whether these
orientations coincide resp. do not coincide. Going over to the 
case at hand, we identify 
$\bC^N$ with the real vector space $\bR^{2N}$ and Euclidean inner product 
$(\,\cdot\, , \,\cdot\,) = \bRe \langle\,\cdot\, , \,\cdot\,\rangle$. 
The tangent space $T_A \bPC^{N-1}$ is canonically identified with
$\{ x \in \bC^N \cong \bR^{2N}| \,\, Ax = 0\}$. 
Each such tangent space may be equipped with the (real) 
$(2N - 2)$-form $*(a \wedge ia)$, defining thus a canonical orientation 
of the complex projective space. Here, $`*'$ denotes the operation of
taking the Hodge dual of an alternating form with help of the above Euclidean inner product.
In the same fashion, the tangent space $T_A E^\perp, EA = 0$ is identified with
$\{ x \in \bC^N \cong \bR^{2N}| \,\, Ax = Ex = 0\}$ and the (real)
$(2N - 4)$-form $*(e \wedge ie \wedge a \wedge ia)$ defines an orientation.
Suppose we are given a surface in projective space parametrized by 
$[\sigma(s_1, s_2)]$. With the above identifications, the tangent space 
at a point $[\sigma_0]$ of this surface is given by the real linear span 
of $x_1, x_2$, where
\begin{eqnarray*}
x_i = \frac{\partial \sigma}{\partial s_i} - \left\langle \sigma_0 \Bigg| 
\frac{\partial \sigma}{\partial s_i} \right\rangle \sigma_0,
\end{eqnarray*}   
and the $2$-form $x_1 \wedge x_2$ defines an orientation. Now let 
$[\sigma_0]$ be also in $E^\perp$, i.e. an intersection point. Then
by definition, $\langle \sigma_0| x_i \rangle = \langle \sigma_0 |
e \rangle = 0$ and one calculates
\begin{eqnarray*}
x_1 \wedge x_2 \wedge *(e \wedge ie \wedge \sigma_0 \wedge i\sigma_0)
= \bIm \left[ \left\langle e \Bigg| \frac{\partial \sigma}{\partial s_1} 
              \right\rangle 
              \left\langle  \frac{\partial \sigma}{\partial s_2} \Bigg| e 
              \right\rangle \right] *(\sigma_0 \wedge i\sigma_0).
\end{eqnarray*}
Hence the sign of this intersection point is given by
\begin{eqnarray}\label{100}
{\rm sign \,\, Im} \left[\left
              \langle e \Bigg| \frac{\partial \sigma}{\partial s_1} 
              \right\rangle 
              \left\langle  \frac{\partial \sigma}{\partial s_2} \Bigg| e 
              \right\rangle \right].
\end{eqnarray}
We are now ready to prove the claims made above about the intersection 
numbers. The maps given in Eqs. $(I), (II), (III)$ define a 
surface in $\bPC^{N-1}$ which is schematically drawn in Fig. \ref{fig2}. The 
arrows define an orientation of this surface and the intersection 
numbers are defined with respect to this orientation and the canonical
orientations given above. 

{\bf Case $\alpha.A$):\,\,} By definition the surface $IA$ is parametrized
by $[\sigma(s, t)]$, where
\begin{eqnarray*}
\sigma(t, s) = (1 - s)\langle \phi(t) | U_{01}\phi_1 \rangle \phi(t) + 
s|\langle \phi(t) | U_{01}\phi_1 \rangle|^2 U_{01}\phi_1, \\
\phi(t) = (1 - t)\langle \phi_2 | U_{23}\phi_3 \rangle U_{012} \phi_2 +
t | \langle \phi_2 | U_{23}\phi_3 \rangle |^2 U_{0123}\phi_3.  
\end{eqnarray*} 
The orientation of this surface is given by the order of its vertices
$P, Q, R$, defined above. It is easily seen that this means that  
this orientation coincides with $\eta \wedge \xi$, where
\begin{eqnarray*}
\eta = \frac{\partial \sigma}{\partial t} - \left\langle \sigma_0 \Bigg| 
\frac{\partial \sigma}{\partial t} \right\rangle \sigma_0,\qquad
\xi = \frac{\partial \sigma}{\partial s} - \left\langle \sigma_0 \Bigg| 
\frac{\partial \sigma}{\partial s} \right\rangle \sigma_0.
\end{eqnarray*}
Now using Eq.~(\ref{100}) and the definition of $\varphi_{IA}$ in
Eq.~(\ref{IA}), we see that the sign of an intersection is given by
\begin{eqnarray*}
{\rm sign\,\,Im} \left[ \frac{\partial \varphi_{IA}(s_2)}{\partial s_2}
\right].
\end{eqnarray*}

{\bf Case $\alpha.B$):\,\,} The surface $IB$ is a triangle, 
parametrized by $[\sigma(s_2, s_3)]$, where 
\begin{eqnarray*}
\sigma(s_2, s_3) = (1 - s_3) \langle V_{\langle 123 \rangle}(s_2)
\phi_3 | \phi_1 \rangle\,\,U_{01} V_{\langle 123 \rangle}(s_2)\phi_3 
+ s_3 |\langle V_{\langle 123 \rangle}(s_2)\phi_3 | \phi_1 \rangle|^2  
\,\,U_{01}\phi_1.
\end{eqnarray*}
With the above notations, $x_2 \wedge x_3$ is the orientation of this 
surface. Setting $e = \phi_0$, using Eq.~(\ref{100}) and the definition of 
$\varphi_{IB}$ in Eq.~(\ref{IB}), it is straightforward to calculate that an intersection 
has sign 
\begin{eqnarray*}
{\rm sign\,\,Im} \left[ \frac{\partial \varphi_{IB}(s_2)}{\partial s_2}
\right].
\end{eqnarray*}

{\bf Case $\beta$):\,\,} Surface $II$ is a square, parametrized
by $[\sigma(s_1, s_3)]$, the orientation of this piece of surface being
given by $x_1 \wedge x_3$.
\begin{eqnarray*}
\sigma(s_1, s_3) = (1 - s_3) \langle \phi_2 | U_{23} \phi_3 \rangle
V_{\langle 012 \rangle}(s_1)\phi_2
+ s_3 |\langle \phi_2 | U_{23} \phi_3 \rangle|^2  
\,\,V_{\langle 012 \rangle}(s_1) U_{23} \phi_3.
\end{eqnarray*}
Now setting $e = \phi_0$, using Eq.~(\ref{100}) and the definition of
$\varphi_{II}$ in Eq.~(\ref{II}), we get the result after a short computation. 

{\bf Case $\gamma$):\,\,} Surface $III$ is a square,  
parametrized by $[\sigma(s_1, s_2)]$, where $x_1 \wedge x_2$ is its
orientation and   
\begin{eqnarray*}
\sigma(s_1, s_2) = V_{\langle 0123 \rangle}(s_1, s_2)\phi_3. 
\end{eqnarray*}
Now setting $e = \phi_0$, it follows immediately from the definition
of $\varphi_{III}$ in Eq.~(\ref{III}) and from Eq.~(\ref{100}) that the 
sign of the intersection is given by 
\begin{eqnarray*}
{\rm sign\,\, Im}
\left[
\frac{\partial \varphi_{III}(s_1, s_2)}{\partial s_1}
\frac{\partial \overline{\varphi_{III}(s_1, s_2)}}{\partial s_2} \right].
\end{eqnarray*}

\end{appendix}

\end{document}